\documentclass[fleqn,usenatbib,letters]{mnras}

\usepackage{newtxtext,newtxmath}

\usepackage[T1]{fontenc}

\DeclareRobustCommand{\VAN}[3]{#2}
\let\VANthebibliography\thebibliography
\def\thebibliography{\DeclareRobustCommand{\VAN}[3]{##3}\VANthebibliography}

\usepackage{graphicx}	
\usepackage{amsmath}	

\DeclareRobustCommand{\ion}[2]{\textup{#1\,\textsc{\lowercase{#2}}}}

\title{Doppler Dimming and Brightening Effects in Solar Prominences}

\author[A. W. Peat, C. M. J. Osborne, and P. Heinzel]{
Aaron W. Peat,$^{1,2}$\thanks{E-mail: aaron.peat@uwr.edu.pl}
Christopher M. J. Osborne,$^{2}$ and
Petr Heinzel$^{1,3}$
\\
$^{1}$University of Wroc\l{}aw, Centre of Scientific Excellence -- Solar and Stellar Activity, Kopernika 11, 51-622 Wroc\l{}aw, Poland\\
$^{2}$SUPA School of Physics and Astronomy, University of Glasgow, Glasgow, G12 8QQ, UK\\
$^{3}$Astronomical Institute, Czech Academy of Sciences, 25165 Ond\v{r}ejov, Czech Republic}
\date{Accepted XXX. Received YYY; in original form ZZZ}

\pubyear{2024} 

\begin{document}
\label{firstpage}
\pagerange{\pageref{firstpage}--\pageref{lastpage}}
\maketitle

\begin{abstract}
We explored the impact that Doppler dimming and brightening effects from bulk motions of solar prominences have on the formation of Ly$~\alpha$, H$~\alpha$, and \ion{Mg}{ii}~h line profiles. We compared two schemes in which these effects manifest; when the prominence is moving radially away from the solar surface (radial case), and when the prominence is moving parallel to the solar surface (horizontal case). To do this, we analysed 13,332 model profiles generated through the use of the 1D NLTE (i.e. departures from Local Thermodynamic equilibrium) radiative transfer (RT) code \textsc{Promweaver}, built on the \textsc{Lightweaver} NLTE RT framework to mimic the behaviour and output of the 1D NLTE RT code PROM. We found that horizontal velocities are just as, or more important than radial velocities. This demonstrates that horizontal velocities need to be accounted for when attempting to do any sort of forward modelling.
\end{abstract}

\begin{keywords}
Sun: filaments, prominence -- radiative transfer -- line: profiles
\end{keywords}



\section{Introduction}

Solar prominences are cool, dense structures suspended in the solar corona which have been observed for centuries \citep[e.g.][]{vassenius_observation_1733, secchi_soleil_1875, hale_spectrohelioscope_1929, vyssotsky_astronomical_1949} and simulated for half a century \citep[e.g.][]{ishizawa_emission_1971, heasley_theoretical_1974,heinzel_formation_1987, gouttebroze_radiative_2004}. It has been long shown that eruptive prominences are affected by Doppler dimming (DDE) and brightening (DBE) effects -- where the incident radiation from the solar disc appears Doppler shifted in the frame-of-reference of the solar prominence, causing a brightening or dimming effect in the scattered line radiation. From here, we will use the acronym DXE to refer to both DDE and DBE. The former occurs when the incident radiation is in emission, and the latter when in absorption \citep{hyder_h_1970,heinzel_hydrogen_1987,labrosse_physics_2010, vial_solar_2015}. 

More generally, when a prominence undergoes some bulk motion, DBE occurs when the incident radiation increases in the wavelength range of the absorption profile of the line in question, and DDE occurs when the incident radiation decreases in the wavelength range of the absorption profile of the line \citep{tandberg-hanssen_nature_1995}.

An underexplored avenue which produces DXE is when a prominence has zero radial velocity, but non-zero velocity parallel to the solar surface \citep[e.g.][]{rompolt_radiation_1967, morimoto_method_2003}. We believe that no large scale quantitative study has been performed to understand the impact of such a velocity field on the radiation observed from solar prominences. In the case of prominences, this is commonly referred to as line-of-sight (LOS) velocity -- however, in observations, true LOS velocities commonly contain some radial component. Here, we wish to demonstrate that, even with a zero radial velocity component, prominences (and filaments) undergoing bulk motions experience significant DXE. 

To begin, we will briefly introduce the NLTE (i.e. departures from Local Thermodynamic equilibrium) radiative transfer (RT) code which we use to simulate solar prominence emission. Following this, we briefly discuss our results and compare them to the effect when caused by non-zero radial velocities. We then discuss the implications this has for NLTE inversions of solar prominence atmospheres. Finally, we offer our conclusions.

\section{Promweaver}
\begin{figure}
    \centering
    \includegraphics[width=\linewidth]{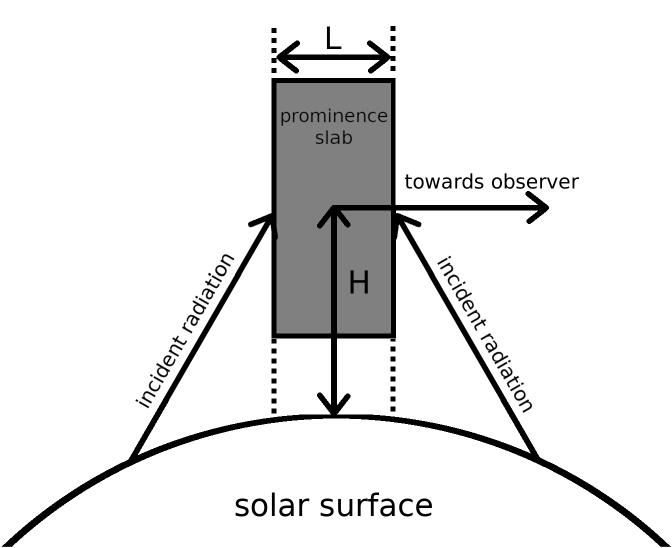}
    \caption{How \textsc{PROM} and \textsc{Promweaver} model a prominence. Where $H$ is the altitude of the prominence, and $L$ is the slab width. This figure is adapted from \citet{peat_diagnostics_2023}.}
    \label{fig:prom}
\end{figure}
\begin{figure*}
    \centering
    \resizebox{\hsize}{!}
    {\includegraphics[]{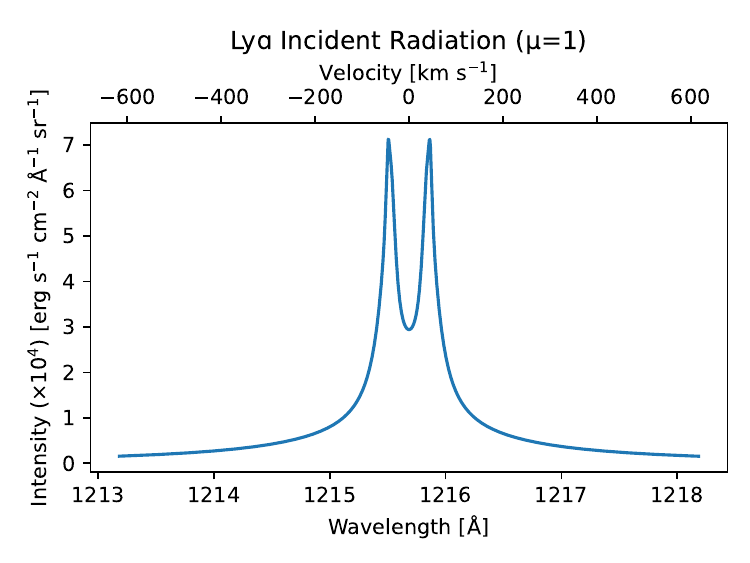}
    \includegraphics[]{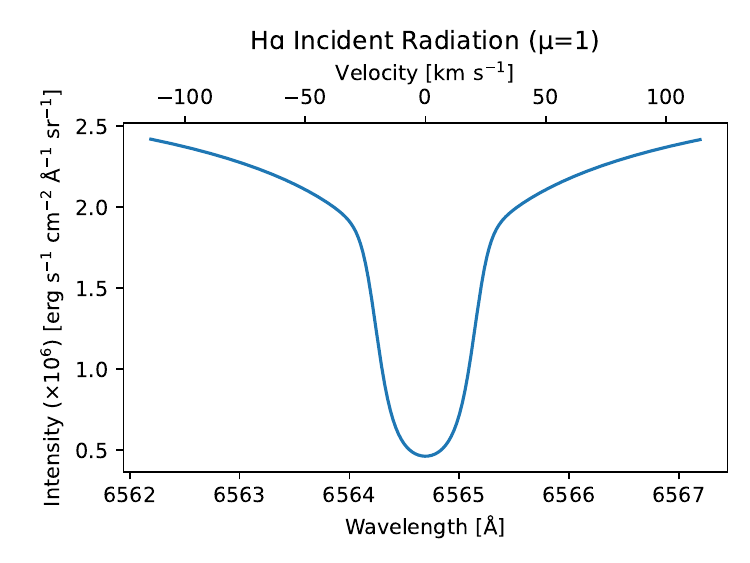}
    \includegraphics[]{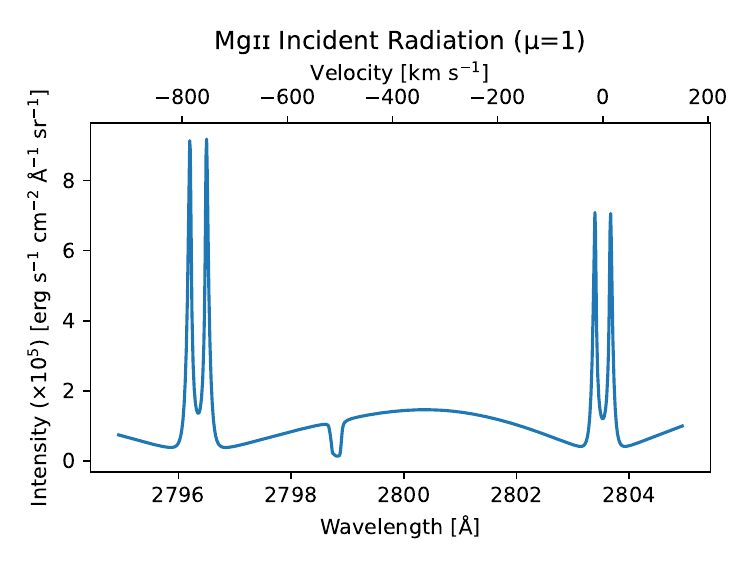}}
    \caption{The incident radiation of Ly~$\alpha$, H~$\alpha$, and \ion{Mg}{ii}~h\&k as seen by the prominence for $\mu=1$ (in the frame of the prominence, i.e. the radiation that would be incident on the prominence from directly below). In the \ion{Mg}{ii}~h\&k panel, the velocities are calculated using \ion{Mg}{ii}~h (2803.53~\AA) as the rest wavelength.}
    \label{fig:inc}
\end{figure*}
To demonstrate how non-zero velocities parallel to the solar surface in solar prominences produce DXE, we employ the use of the 1D NLTE RT code \textsc{Promweaver} \citep{promweaver}, built on top of the NLTE RT framework, \textsc{Lightweaver} \citep{osborne_lightweaver_2021}. \textsc{Promweaver} is written to efficiently and expediently replicate the results of the 1D NLTE RT code \textsc{PROM} \citep{gouttebroze_hydrogen_1993,heinzel_theoretical_1994, labrosse_nonlte_2004, levens_modelling_2019}. However, unlike \textsc{PROM}, \textsc{Promweaver} has the ability to include any arbitrary bulk velocity field, meanwhile \textsc{PROM} can only include radial velocities. \textsc{Promweaver} is not the first prominence code that can include arbitrary velocities; for example, \textsc{MALI} \citep{heinzel_multilevel_1995, heinzel_formation_2014} includes an implementation of the technique discussed in \citet{heinzel_hydrogen_1987}. However, in practice, it has only been used for radial velocities in the context of erupting prominences. Additionally, PROM produces half line profiles which are mirrored to produce the full profiles, but \textsc{Promweaver} synthesises the entire line profile handling arbitrary velocity fields and interactions between overlapping transitions.

\textsc{Promweaver} allows us to simulate any of the commonly modelled and observed lines; here we focus on Ly~$\alpha$, H~$\alpha$, and \ion{Mg}{ii}~h {using a 5-level plus continuum H model and a 10-level plus continuum Mg model, following \citet{leenaarts_formation_2013-1}}.
Partial frequency redistribution (PRD) is of critical importance to the formation of the Ly~$\alpha$ line \citep{heinzel_formation_1987}. Additionally, PRD plays a key role in the formation of the \ion{Mg}{ii} resonance lines \citep{paletou_two-dimensional_1993, heinzel_formation_2014}, but this only affects the far wings of the emission, which may be below the detection limit \citep{heinzel_formation_2014}. Here we treat all resonance lines in PRD. {These atoms are solved simultaneously, allowing for mutual radiative interactions to occur.}

\textsc{Promweaver}, like \textsc{PROM} or \textsc{MALI}, models a solar prominence as a monolithic slab suspended in the solar corona (see Figure \ref{fig:prom}). 
{It solves for statistical equilibrium of the considered atoms in the given slab, additionally conserving charge, and modifying the mass density of the structure to follow the prescribed pressure stratification (which is affected by ionisation).}
\textsc{Promweaver} can use one of two model atmospheres; isothermal and isobaric, and one with a prominence-to-corona transition region (PCTR). The PCTR is the region where the cool and dense prominence transitions into the hot and tenuous corona. This region is {modelled} by the following equations \citep{kippenhahn_eine_1957, anzer_energy_1999, labrosse_nonlte_2004},
\begin{equation} 
    T(m)=T_{\text{cen}}+(T_{\text{tr}}-T_{\text{cen}})\left(1-4\frac{m}{M}\left(1-\frac{m}{M}\right)\right)^\gamma,
    \label{tstrat}
\end{equation}
\begin{equation}
    p(m)=4p_c\frac{m}{M}\left(1-\frac{m}{M}\right)+p_{\text{tr}},
    \label{pstrat}
\end{equation}
for $\gamma\geq2$, and where $p_c=p_{\text{cen}}-p_{\text{tr}}$. Here $m$ is the column mass, $M$ is the total column mass, $T(m)$ is the temperature as a function of column mass, $T_{\text{cen}}$ is the central (prominence) temperature of the plasma, $T_{\text{tr}}$ is the (coronal) temperature at the edge of the PCTR, and $\gamma$ is a dimensionless factor that defines the extent of the PCTR. $p(m)$ is the pressure as a function of column mass, $p_{\text{cen}}$ is the central (prominence) pressure, and $p_{\text{tr}}$ is the (coronal) pressure at the edge of the PCTR. Eq. \ref{tstrat} is an empirical expression described in \citet{anzer_energy_1999}, while Eq. \ref{pstrat} is an analytical formula from the Kippehahn-Schl\"{u}ter model \citep[KS;][]{kippenhahn_eine_1957, heasley_structure_1976} for magnetohydrostatic (MHS) equilibrium. These equations are used as \textsc{Promweaver} is built to mimic PROM and this is how the PCTR is constructed in \textsc{PROM} \citep{labrosse_nonlte_2004}. As Eq. \ref{pstrat} comes from MHS equilibrium, it should be stressed that the applicability of this model for erupting prominences is questionable.

The incident radiation is generated using the FAL C atmosphere \citep{falc93} and `light cone' method discussed in the appendix of \citet{jenkins_non_2023}. The incident radiation at $\mu=1$ (i.e. radially from the solar surface) in the lines considered is shown in Fig. \ref{fig:inc}. In all situations we employ fully angularly dependent boundary conditions including the effects of limb darkening (self-consistently in a plane-parallel approximation as per \citet{jenkins_non_2023}) and arbitrary prominence motions.

{The emergent intensity is computed by a final pass through the model using the computed level populations and ratio of the PRD emission to absorption profiles.
The cones in the boundary conditions are disabled to compute the radiation along a specific ray (rather than the spherical integrals used to solve the non-LTE problem).
This method can handle arbitrary viewing angles through the prominence slab, including when it is partially back-lit against the solar disc.}
 
\begin{table}
\centering
\begin{tabular}{cccc}
\hline\hline
Parameter  & Unit     & Iso Models & PCTR Models \\ \hline
$T_{\text{cen}}$       & kK       & 8         & 6          \\
$T_{\text{tr}}$        & kK       & --       & 100        \\
$p_{\text{cen}}$       & dyn~cm$^{-2}$ & 0.05      & 0.1        \\
$p_{\text{tr}}$        & dyn~cm$^{-2}$ & --       & 0.01       \\
L & km       & 400      &   407.69$^a$       \\
M          & g~cm$^{-2}$   &    2$\times10^{-6}$ $^a$      & 5$\times10^{-6}$       \\
$v_T$         & km~s$^{-1}$    & 5         & 5          \\
$\gamma$      &         & --       & 4          \\ \hline\hline
\end{tabular}
\caption{The base model parameters. For the isothermal and isobaric (iso) models, $T_{\text{cen}}$ and $p_{\text{cen}}$ are taken as $T$ and $p$, respectively. The iso models are built using the slab width, while the PCTR models use maximum column mass to limit the extent of the structure.\\$^a$ For this model type, this parameter is not strictly an input. It instead is found when solving statistical equilibrium.}
\label{table:params}
\end{table}

Using \textsc{Promweaver}, we generated a large grid of models. These models are all based on one of two base models, which can be seen in Table \ref{table:params}. For isothermal models, slab width (L) is an input and the mass density (M) is calculated when solving statistical equilibrium. Meanwhile, for the PCTR models, mass density is an input, and the slab width is calculated during the solving of statistical equilibrium. Due to these differences, we attempted to select variables such that the slab widths and mass densities of these model types were of similar orders of magnitudes. To investigate the effect of bulk motion on the line profiles, we create a modified set of these base models such that they have horizontal velocities in the range 0--80~km~s$^{-1}$ with a step size of 0.8~km~s$^{-1}$. Another set was also created with radial velocities in the same range in order to directly compare the two effects. As DXE is affected by altitude \citep{rompolt_radiation_1967, rompolt_ha_1980, rompolt_ba_1980, rompolt_lya_1980}, we further modified these models to include altitudes in the range 5--70~Mm with a step size of 2~Mm. These combinations resulted in a total of 13,332 models.

\begin{figure*}
    \centering
    \resizebox{\hsize}{!}
    {\includegraphics[]{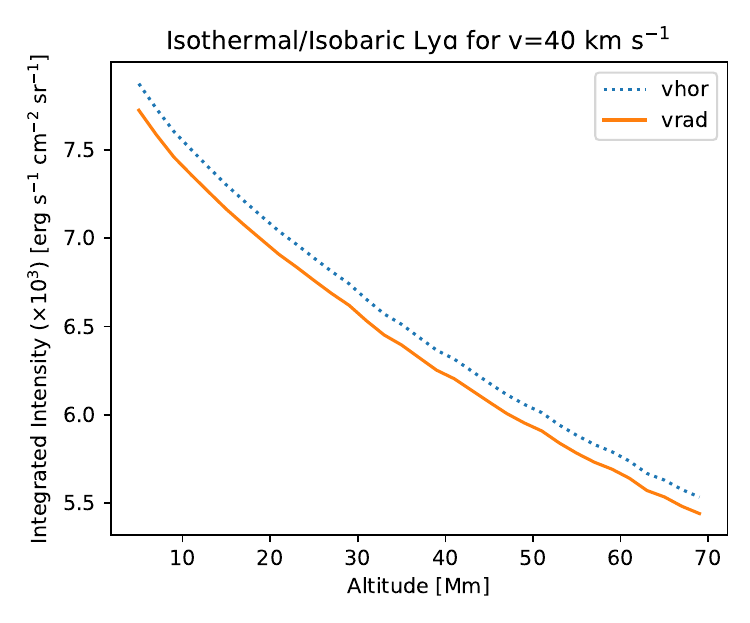}
    \includegraphics[]{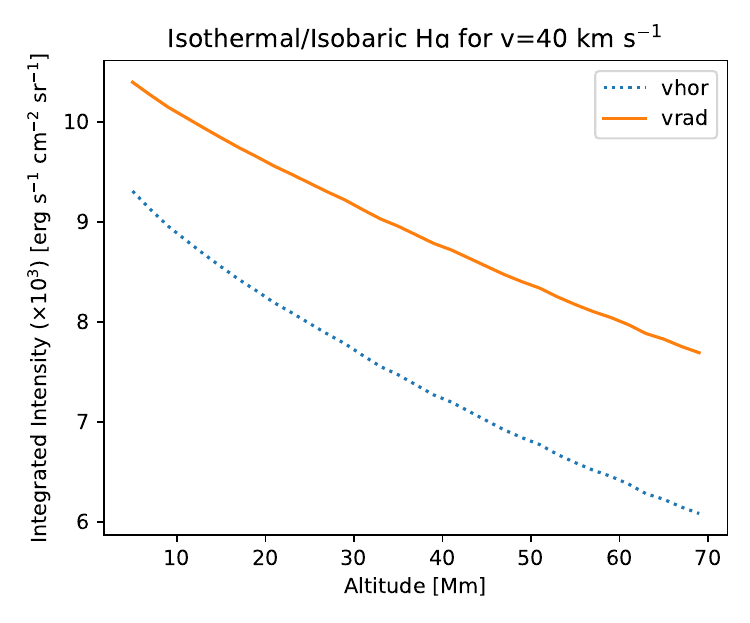}
    \includegraphics[]{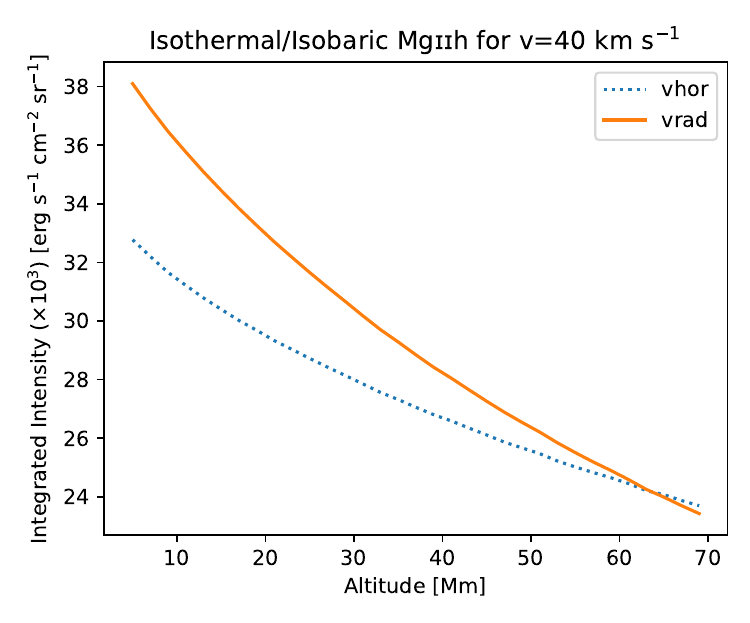}}
    \resizebox{\hsize}{!}
    {\includegraphics[]{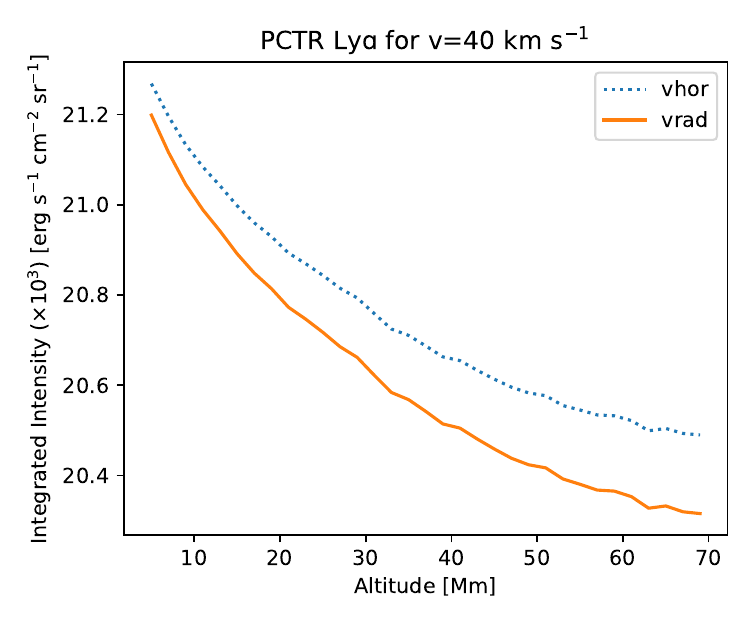}
    \includegraphics[]{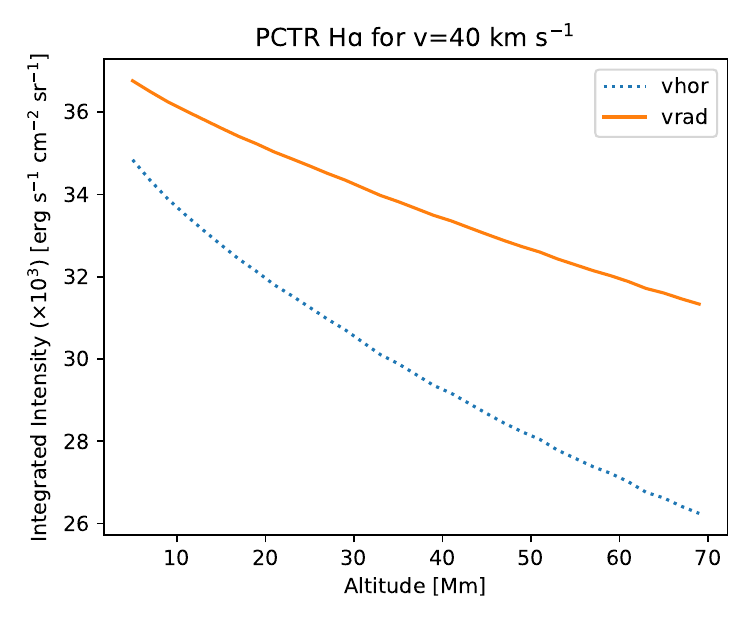}
    \includegraphics[]{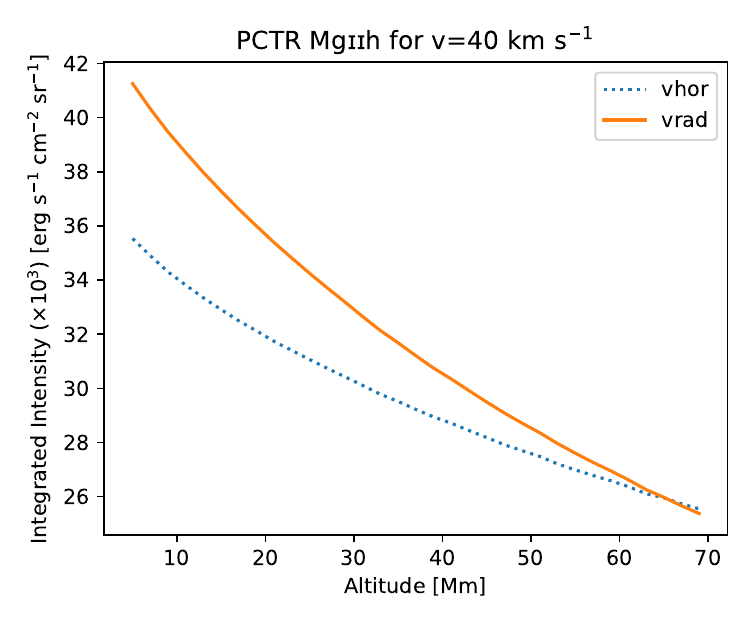}}
    \caption{How DD/DB effects from horizontal (vhor) and radial velocities (vrad) are affected by a change in height. The top row is the results from the isothermal and isobaric atmospheres, and the bottom the PCTR atmospheres.}
    \label{fig:fixvel}
    \centering
    \resizebox{\hsize}{!}
    {\includegraphics[]{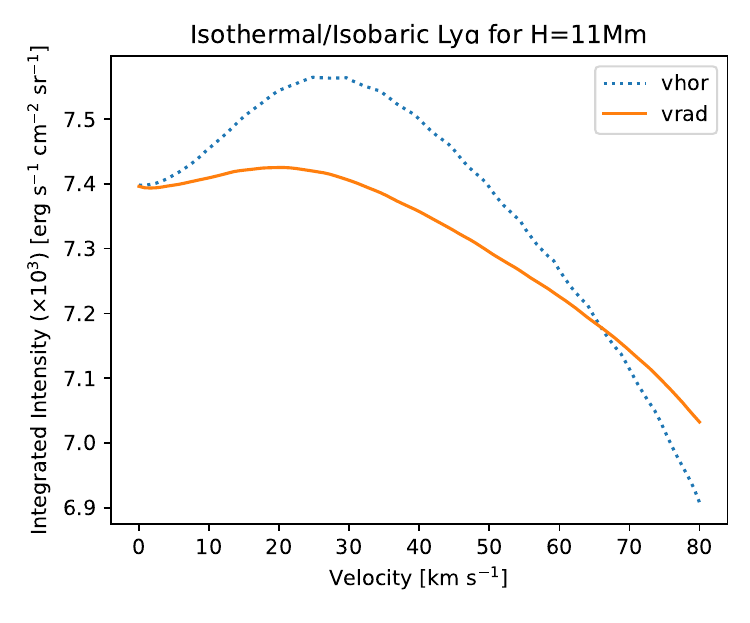}
    \includegraphics[]{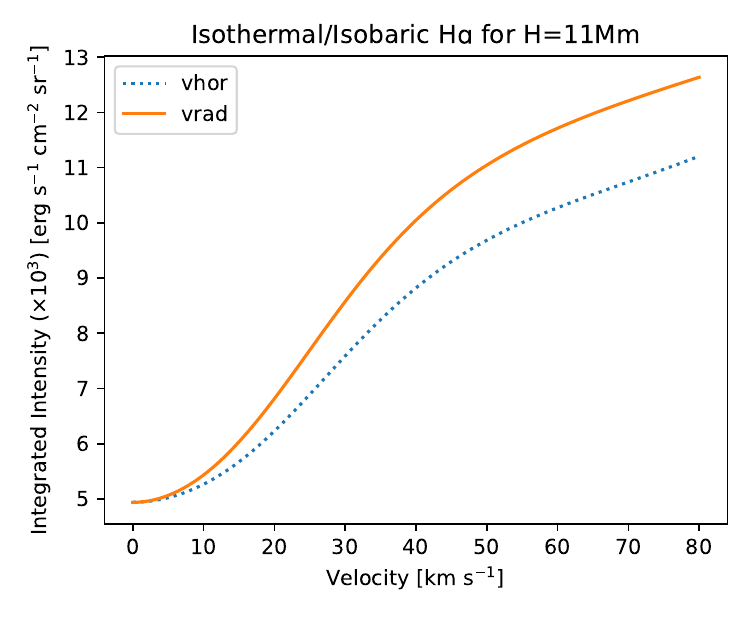}
    \includegraphics[]{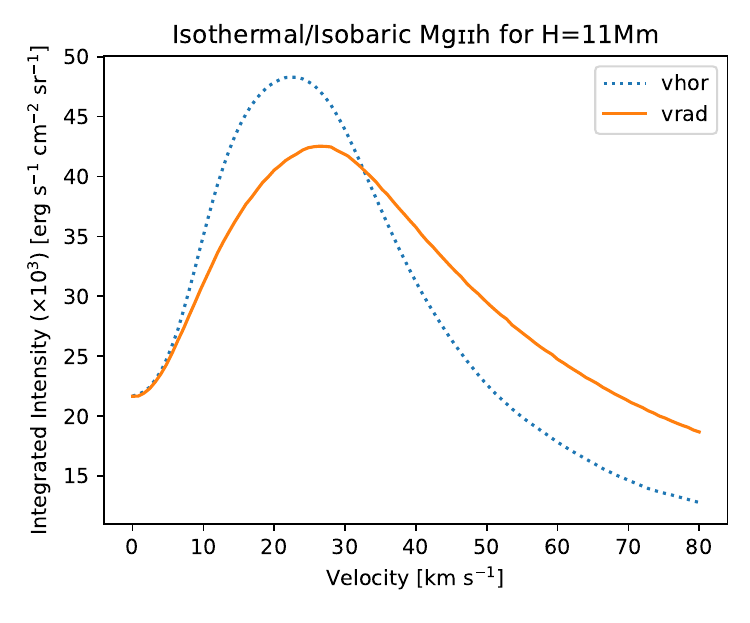}}
    \resizebox{\hsize}{!}
    {\includegraphics[]{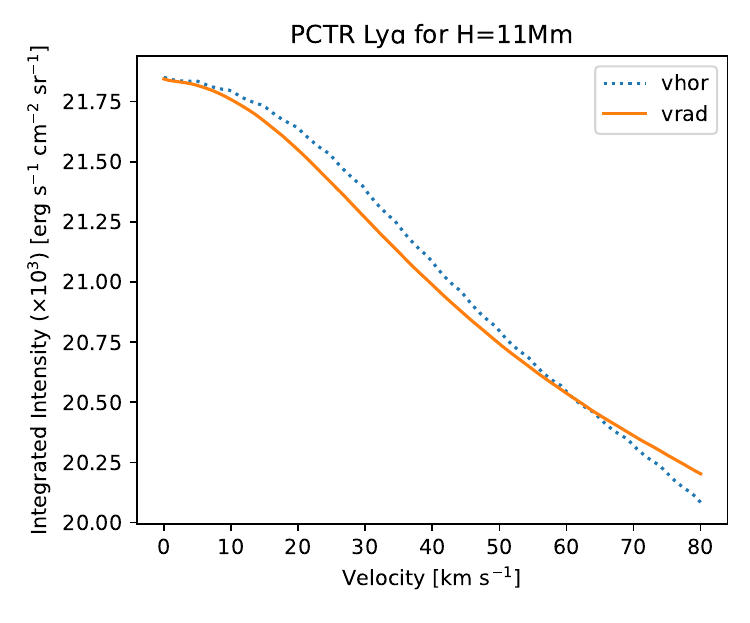}
    \includegraphics[]{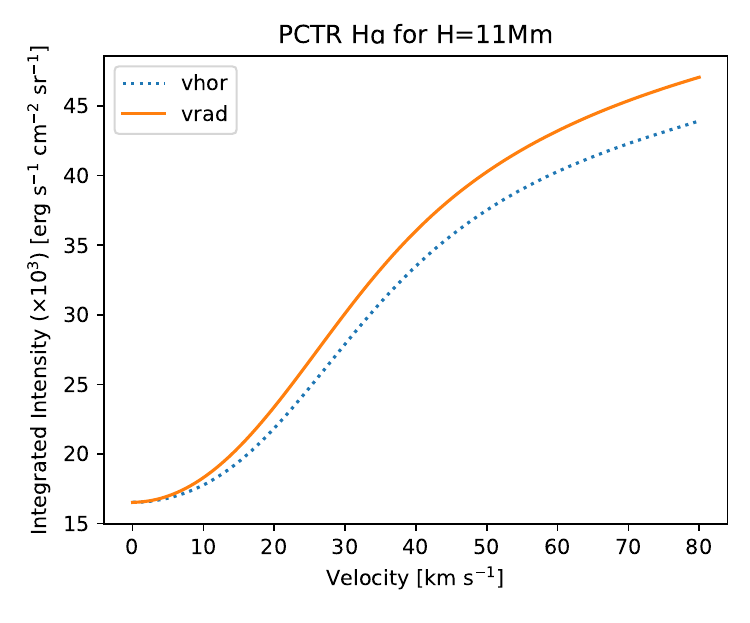}
    \includegraphics[]{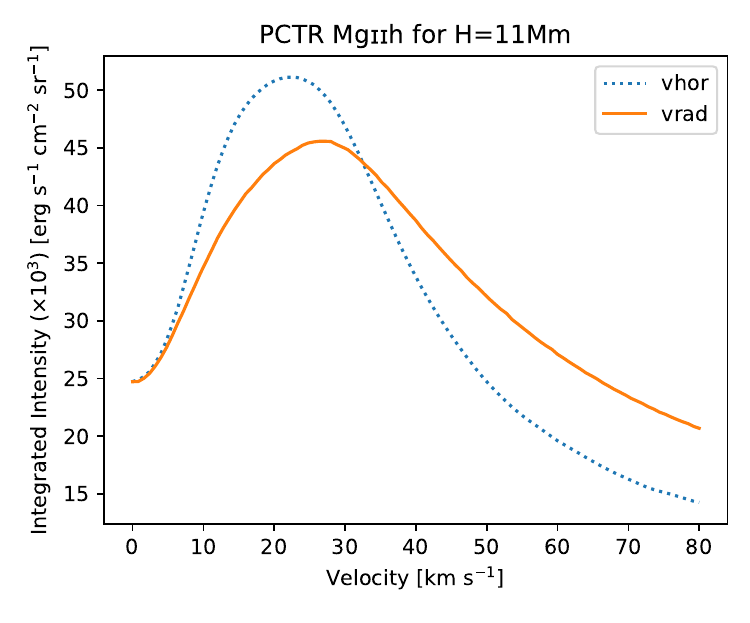}}
    \caption{How DD/DB effects from horizontal (vhor) and radial velocities (vrad) are affected by a change in velocity. The top row is the results from the isothermal and isobaric atmospheres, and the bottom the PCTR atmospheres.}
    \label{fig:fixh}
\end{figure*}

These models were generated on a 36-core/72-thread dual socket motherboard machine with two Intel Xeon Gold 6140s with a base clock speed of 2.30~GHz, and a turbo speed of 3.70~GHz. The wall time\footnote{Real world time (i.e. measured on a clock on the wall).} taken to produce the 13,332 models was approximately 70 hours with a CPU time\footnote{The total time contributed by all CPU threads.} of approximately 5,000 hours.

\section{Results}

DXE is effectively produced when a prominence undergoes some sort of bulk motion, causing the incident radiation to appear Doppler shifted in the frame of the prominence. For incident radiation seen in absorption, this causes a DBE as the shifted wings of the incident line profile increase the radiation scattered by the prominence. For incident radiation seen in emission, the wings also illuminate the prominence, but since the incident wings are dimmer than the incident line core, the net incident radiation is decreased and a DDE is observed. With double peaked emission profiles, at low velocities, we expect to see a DBE as the horns of the incident radiation are now impacting the outgoing radiation. At higher velocities, the wings of incident line profiles take precedence resulting in a DDE.

All 13,332 of our {statistical equilibrium calculations} successfully converged. The general relationships between velocity and height for the different model atmospheres and velocity types can be seen in Figs. \ref{fig:fixvel} and \ref{fig:fixh}. A value of 40~km~s$^{-1}$ was chosen as the fixed velocity for Fig. \ref{fig:fixvel} simply because it was the middle of our velocity grid. A value of 11~Mm was chosen as the fixed altitude for Fig. \ref{fig:fixh} as altitudes of the order of 10~Mm are commonly used in papers which use \textsc{PROM} \citep[e.g.][]{zhang_launch_2019, levens_modelling_2019, ruan_diagnostics_2019, peat_solar_2021}. Due to our step size, however, the closest altitude to 10~Mm was 11~Mm.

Fig. \ref{fig:fixvel} shows how the {wavelength} integrated intensity of the emission changes with altitude. The irregularities in the Ly~$\alpha$ curves here are simply due to numerical errors {because of a slight misalignment between the wavelength grids of the incident radiation and the emitted radiation from the prominence}.
Unsurprisingly, as the height increases, the integrated intensity falls -- the Sun presents a smaller solid angle to the prominence. As the height increases the {radial velocity curves were expected to produce a greater DXE effect than in the horizontal case. This is due to the lessening of the horizontal component of the incident rays as height increases. However, \ion{Mg}{ii}~h does not exhibit this behaviour until higher altitudes.} H~$\alpha$ and Ly~$\alpha$ both appear to diverge instead, with the two curves for Ly~$\alpha$ in an isothermal and isobaric atmosphere appearing essentially parallel. The integrated intensities of the lines in PCTR atmospheres are almost three times higher than in isothermal and isobaric atmospheres for the Hydrogen lines, and approximately only 10~per cent higher in \ion{Mg}{ii}~h.

Fig. \ref{fig:fixh} shows how the integrated intensity of the emission changes with velocity for a fixed altitude. In H~$\alpha$, for both atmosphere types, radial and horizontal velocities produce a similar brightening effect in both behaviour and value. 
For Ly~$\alpha$ and \ion{Mg}{ii} however, we see much more complicated behaviour. Up to 30~km~s$^{-1}$, horizontal velocities produce a much larger DBE in \ion{Mg}{ii}~h than the radial velocities. At higher velocities, the horizontal velocities also produce a much greater DDE than radial. The behaviour of \ion{Mg}{ii} in the radial case is consistent with that found by \cite{heinzel_formation_2014}.
Ly~$\alpha$ sees a similar effect, but only in the isothermal and isobaric atmosphere with the horizontal velocity dimming at a much greater rate than the radial velocity after its peak brightening. In the PCTR atmosphere, there is a similar, but much less pronounced, relationship.  

The asymmetry of the lines was briefly investigated. \citet{gontikakis_emission_1997} showed that when treated in PRD, the Ly~$\alpha$ line profiles from moving solar prominences are asymmetric. This is not possible under CRD, and another reason why we treated all of the resonance lines in PRD.  Here, the asymmetry was calculated via the quantile method \citep{kerr_iris_2015}. This works by calculating the cumulative distribution function (CDF) of a line profile, and normalising them such that the values in the function range from 0--1. The wavelength at which the CDF is 0.5 ($\lambda_{50}$) is defined as the line core. If the line is of a Gaussian shape, the difference between $\lambda_{88}$ and $\lambda_{12}$ is the full width half maximum. From this, we calculate the asymmetry by the following relationship,
\begin{equation}
    \text{Asymmetry}=\frac{\left(\lambda_{88}-\lambda_{50}\right)-\left(\lambda_{50}-\lambda_{12}\right)}{\lambda_{88}-\lambda_{12}}=\frac{\lambda_{88}-2\lambda_{50}+\lambda_{12}}{\lambda_{88}-\lambda_{12}},
\end{equation}
where a positive asymmetry means there is more emission in the blue, and a negative asymmetry means there is more emission in the red. The asymmetry in all of the line profiles is not appreciably affected in either the horizontal or radial velocity modes, never exceeding $\pm$0.02. However, in observations, asymmetry is usually measured on the order of $\pm$0.5 \citep{ruan_dynamic_2018, peat_solar_2021}. Therefore, this effect would be negligible in observations and cannot be a source of the asymmetry on its own.

\section{Discussion}

The way in which DXE occurs in the radial and horizontal velocity cases are subtly different. In the much studied radial case, as the prominence moves away from the solar surface, the incident radiation appears red-shifted in the frame of the prominence, and if it is moving towards the the solar surface, the radiation appears blue-shifted. However, for the horizontal case, the leading edge of the prominence sees the incident radiation as blue-shifted, while the trailing edge sees the incident radiation as red-shifted. The radial case is only affected by one side of the wings of the incident radiation, whereas the horizontal case is affected by both sides of the wings. 

The shape and intensity of the incident radiation profiles is extremely important to the total intensity of the line profiles produced by a prominence whether it is undergoing some bulk motion or not. Here we used the Ly~$\alpha$, H~$\alpha$, and \ion{Mg}{ii}~h emission from the aforementioned FALC atmospheric model as the incident radiation. Whereas other studies modelling the emission from prominences use observations of the lines in question \citep[e.g.][]{heinzel_hydrogen_1987, levens_modelling_2019}. However, if the incident radiation were to change from FALC to that found in observations, the qualitative relationships between the intensities and velocities would remain the same.

Much of the focus of DXE has been contextualised in the form of eruptive prominences moving predominantly radially \citep[e.g.][]{heinzel_hydrogen_1987,vial_solar_2015, zhang_launch_2019}, but this work shows that horizontal velocities also play a role. In observation, eruptive prominences are never going to erupt perfectly perpendicular to the solar surface. This will introduce a small horizontal component to its velocity, which is when the horizontal effect is at its greatest for lines with incident radiation in emission (see Fig. \ref{fig:fixh}). Fig. \ref{fig:fixvel} shows that radial and horizontal DXE appear to dissipate with height at the same rate, illustrating that this horizontal component will continue to be important as the prominence moves away from the solar disc.
Furthermore, it is likely that the solar spectra incident on an erupting prominence are no longer symmetrical, increasing the complexity of the DXE effect.
If it is assumed that an erupting prominence only has radial velocity, the radial velocity will be overestimated to account for the DXE introduced by the horizontal effect.

\section{Conclusions}

In this study we demonstrated an application of the NLTE \textsc{Lightweaver} RT framework, in the context of \textsc{Promweaver}, to explore DXE caused by different orthogonal motions. To meet this goal, we generated 13,332 models of differing parameters.

We showed the importance of all translational bulk motions of solar prominences on the formation of observed line profiles. Past studies mainly focused on radial velocities to better understand erupting prominences, but neglected horizontal velocities entirely. This neglect may have caused derived radial velocities of eruptive prominences from forward modelling to be overestimated. 

In both the isboaric/isothermal and PCTR atmospheres, Ly~$\alpha$ and \ion{Mg}{ii}~h, as velocity increases, the horizontal case produces a much stronger DXE effect than its radial counterpart, but H~$\alpha$ sees the greater effect caused by the radial case. With respect to increasing altitude, the intgrated intensities in the radial and horizontal cases both appear to diminish at equivalent rates in the two hydrogen lines. However, with respect to \ion{Mg}{ii}~h, the radial and horizontal cases converge. Perhaps at larger velocities, Ly~$\alpha$ and H~$\alpha$ will also display this behaviour. In general, the magnitude of the DXE produced by the horizontal case is comparable to that produced in the radial case. In future studies, care should be taken when forward modelling to account for effects introduced by horizontal velocities and not to erroneously attribute it to only radial velocities.

\section*{Acknowledgements}

AWP and PH acknowledge financial support from the `Excellence Initiative--Research University' program for the years 2020--2026 at the University of Wroc\l{}aw, project No. BPIDUB.4610.96.2021.KG. AWP would also like to thank Dr. Nicolas Labrosse for having the time for many impromptu conversations.

CMJO is grateful for the support of the Royal Astronomical Society's Norman Lockyer fellowship, and the support of the University of Glasgow's College of Science and Engineering and Lord Kelvin/Adam Smith fellowship.

PH was also supported by the grant No. 22-34841S of the Czech Funding Agency. He would like to express his deep gratitude to Prof. Bogdan Rompolt from the University of Wroc\l{}aw for introducing him to problems
of DXE.

This research used version 3.8.2 of Matplotlib \citep{matplotlib}, version 1.24.4 of NumPy \citep{harris_array_2020}, and version 5.3.3 of Astropy (\href{http://www.astropy.org}{http://www.astropy.org}) a community-developed core Python package for Astronomy \citep{the_astropy_collaboration_astropy_2013, the_astropy_collaboration_astropy_2018}.

\section*{Data Availability}
This research presented an application of version 0.1.1 of \textsc{Promweaver} available at \href{https://doi.org/10.5281/zenodo.6546897}{https://doi.org/10.5281/zenodo.6546897}. The most recent version of \textsc{Promweaver} is available to download via \textsc{pip} or at \href{https://github.com/Goobley/Promweaver}{https://github.com/Goobley/Promweaver}. \\
All of the models used in this study can be found in the form of a Python pickle at \href{https://doi.org/10.5281/zenodo.12083238}{https://doi.org/10.5281/zenodo.12083238}.



\bibliographystyle{mnras}
\bibliography{ref} 








\bsp	
\label{lastpage}
\end{document}